\documentclass[english,american,longbibliography,twocolumn,flushbottom,prl]{revtex4-2}
\usepackage[T1]{fontenc}
\usepackage[utf8]{inputenc}
\usepackage{xcolor}
\usepackage{babel}
\usepackage{array}
\usepackage{verbatim}
\usepackage{cprotect}
\usepackage{booktabs}
\usepackage{multirow}
\usepackage{varwidth}
\usepackage{amsmath}
\usepackage{amssymb}
\usepackage{braket}
\usepackage{graphicx}
\usepackage[pdfusetitle, bookmarks=true,bookmarksnumbered=false,bookmarksopen=false, breaklinks=false,pdfborder={0 0 1},backref=false,colorlinks=true]{hyperref}
\hypersetup{
linkcolor=magenta,urlcolor=blue,citecolor=blue,pdfstartview={FitH},hyperfootnotes=false,bookmarks=False}
\makeatletter

\date{\today}
\allowdisplaybreaks

\renewcommand{\big}{\bBigg@\@ne}
\renewcommand{\Big}{\bBigg@{1.5}}
\renewcommand{\bigg}{\bBigg@\tw@}
\renewcommand{\Bigg}{\bBigg@{2.5}}

\newcommand{\biggg}{\bBigg@\thr@@}
\newcommand{\Biggg}{\bBigg@{3.5}}

\def\CRA{CeRh$_{2}$As$_{2}$}

\makeatother

\newcommand{\fock}[1]{\mathrm{#1}}
\newcommand{\Hloc}{\fock{H}_{\mathrm{loc}}}
\newcommand{\Fq}[1]{\fock{F}_{#1}}
\newcommand{\Cp}[1]{\fock{C}_{#1}}
\newcommand{\Phib}[1]{\Phi^{[#1]}}
\newcommand{\phib}[1]{\phi^{[#1]}}
\newcommand{\DCEF}{\Delta_{\mathrm{CEF}}}
\newcommand{\dFE}{\Delta\mathcal{F}}
\newcommand{\Tr}{\mathrm{Tr}}

\begin{document}

\title{Multiorbital periodic Anderson model for \CRA}

\author{Nico A. Hackner}
\email{hackner.nico@gmail.com}
\affiliation{Department of Physics and the MacDiarmid Institute for Advanced Materials
and Nanotechnology, University of Otago, P.O. Box 56, Dunedin 9054,
New Zealand}
\author{Changhee Lee}
\affiliation{Department of Physics and the MacDiarmid Institute for Advanced Materials
and Nanotechnology, University of Otago, P.O. Box 56, Dunedin 9054,
New Zealand}
\author{P. M. R. Brydon}
\email{philip.brydon@otago.ac.nz}
\affiliation{Department of Physics and the MacDiarmid Institute for Advanced Materials
and Nanotechnology, University of Otago, P.O. Box 56, Dunedin 9054,
New Zealand}

\begin{abstract}
CeRh$_2$As$_2$ is a heavy-fermion superconductor with distinct low- and high-field superconducting phases. The phase diagram of this compound also contains other ordered phases, whose nature and origin remain debated. To shed light upon these states, we
develop a periodic Anderson model for CeRh$_2$As$_2$ in which the $\Gamma_7$--$\Gamma_6$ quasi-quartet of the Ce $4f$ electrons hybridizes with Rh $d$ electrons. The onsite $4f$ interactions are treated within rotationally invariant slave-boson mean-field theory, with double occupancy excluded. In the paramagnetic state, the single-orbital model widely used in the literature emerges as a low-energy limit of our model, with its spin-orbit coupling arising from the $f$--$d$ hybridization together with local inversion-symmetry breaking. Allowing for commensurate magnetic order, we find that the ordering wave vector is set mainly by the electron filling. In-plane antiferromagnetic order with $\mathbf{Q}=(\pi,\pi)$ is favored at low filling, and A-type antiferromagnetic order with $\mathbf{Q}=(0,0)$ and $c$-axis moments, in close competition with ferromagnetism, at high filling. Induced quadrupole moments are generally small, becoming substantial only when the $\Gamma_7$ and $\Gamma_6$ doublets are nearly degenerate.
\end{abstract}

\maketitle

\section{Introduction}\label{sec:intro}

The superconductivity of the heavy-fermion compound CeRh$_2$As$_2$ is remarkable: For fields close to the $c$ axis, a field-induced transition between distinct low- and high-field pairing states is observed, with the high-field phase persisting far beyond the Pauli limit \cite{Khim2021,Khanenko2025}. This behavior has been widely understood as reflecting a switch in the parity of the superconducting order parameter \cite{Khim2021,Landaeta2022, Khanenko2025,Schertenleib2021,Mockli2021,Cavanagh2022,Nogaki2022,Ogata2023}. Specifically, the nonsymmorphic crystal structure of CeRh$_2$As$_2$ implies a sublattice degree of freedom, with Rashba-like spin-orbit coupling allowed by the locally broken inversion symmetry on each sublattice. Assuming that pairing is predominantly intrasublattice and that the coupling between the sublattices is small compared to the spin-orbit coupling, pairing states with the same (even parity) and opposite (odd parity) signs on each sublattice are nearly degenerate. An applied Zeeman field suppresses the even-parity state, allowing the appearance of the odd-parity state.

Despite its success, this scenario is based on a highly simplified description of the electronic structure of CeRh$_2$As$_2$. In particular, it does not account for the Kondo effect or the strong electronic correlations associated with the Ce 4$f$ electrons. Moreover, the crystal-electric-field splitting of the $J=5/2$ multiplet in this compound is unusual, with the lowest $\Gamma_7$ doublet lying only about 30 K below the next-lowest $\Gamma_6$ doublet; this splitting happens to coincide with the estimated Kondo coherence temperature \cite{Hafner2022,Christovam2024}. This so-called ``quasi-quartet'' structure suggests that both doublets may be important for the low-energy physics and has motivated several authors to identify quadrupolar degrees of freedom as playing an important role in this compound \cite{Hafner2022,Christovam2024,Semeniuk2026,Schmidt2024,Thalmeier2025,Numa2026}. Nevertheless, the quasi-quartet structure is absent from most theoretical descriptions of the superconductivity, which retain only the lowest crystal-field doublet \cite{Schertenleib2021,Mockli2021,Cavanagh2022,Nogaki2022}.

A further complication is the presence of other types of order in the phase diagram. Specifically, the superconductivity at zero field occurs within the so-called ``phase I,'' which is believed to be predominantly antiferromagnetic \cite{Kibune2022,Chajewski2024Expt,TChen2024Expt,Ogata2024,Khim2025uSR,Galeski2026,Ogata2026}, although the ordering vector is controversial. While the interplay of phase I with superconductivity can be reasonably understood within a Ginzburg--Landau description \cite{CLee2026PD,Jakubczyk2025}, the microscopic origin and nature of this order remain unclear. Intriguingly, phase I undergoes a first-order transition into a distinct ``phase II'' under in-plane magnetic field, whose critical temperature grows with the field strength \cite{Hafner2022,Semeniuk2026}. Such behavior is unusual for purely antiferromagnetic order and has therefore motivated proposals involving quadrupolar degrees of freedom.

Theoretical attempts to account for the magnetic and quadrupolar physics have generally adopted a picture of localized Ce 4$f$ electrons \cite{Schmidt2024,Thalmeier2025,Numa2026}. Such approaches have had some success in explaining the transition between phases I and II. Nevertheless, they are difficult to reconcile with the observed heavy-fermion character of CeRh$_2$As$_2$, in which the 4$f$ electrons become itinerant through hybridization with the conduction electrons \cite{XChen2024PRX,BChen2024Expt,Wu2024Expt}. Models based on an itinerant description \cite{CLee2025RG,CLee2026} have instead suggested the presence of strong incommensurate spin fluctuations that may be the origin of the ordered phases. However, these results were obtained within weak-coupling approximations, which do not adequately capture the strong correlations of the heavy quasiparticles arising from the underlying intra-atomic interactions. 

To obtain an itinerant description while capturing the atomic physics, here we propose a two-dimensional periodic Anderson model for CeRh$_2$As$_2$ which includes the hybridization of the quasi-quartet levels with itinerant $d$ electrons originating on the Rh sites \cite{Ptok2021,Nogaki2021,Cavanagh2022,Ishizuka2023,Nogaki2026}. Using the rotationally invariant slave-boson (RISB) method \cite{Lechermann2007}, we enforce a condition of at most single occupancy on the Ce sites. We perform a mean-field analysis of the RISB Hamiltonian, which allows us to calculate the renormalized band structure and search for a ground state with commensurate magnetic order. We first study the paramagnetic state in Sec.~\ref{sec:para}. Here we show that the single-orbital tight-binding model of previous theoretical treatments emerges as a limiting case of our model, valid when the conduction band is well separated from the $f$ levels. Our construction thus provides a microscopic origin for the minimal model, in particular its spin-orbit coupling, which arises from the $f$--$d$ hybridization together with local inversion-symmetry breaking. The magnetic phase diagram is examined in Sec.~\ref{sec:magnetic}. We find that it separates into two distinct regimes: in-plane antiferromagnetism at $\mathbf{Q}=(\pi,\pi)$ at low filling, and A-type antiferromagnetic or ferromagnetic order at $\mathbf{Q}=(0,0)$ at high filling.
Although quadrupole moments are present in most of the ordered states, they are typically small. They can nevertheless become significant upon reducing the separation of the $\Gamma_7$ and $\Gamma_6$ levels. We conclude in Sec.~\ref{sec:discussion} with a discussion of the implications for our understanding of CeRh$_2$As$_2$.

\section{Periodic Anderson model}\label{sec:model}

We construct our periodic Anderson model (PAM) by including $J=5/2$ $f$-electron states localized on the Ce atoms and conduction $d$ electrons originating on the Rh atoms \cite{Ptok2021,Cavanagh2022} and accounting for their nonlocal hybridization through the Slater--Koster two-center approximation \cite{SlaterKoster1954,Sharma1979,Takegahara1980}. The Ce ions occupy Wyckoff position $2c=(\tfrac12 \mp \tfrac14,\tfrac12 \mp \tfrac14,\pm z)$ with $z=0.75469$~\cite{Khim2021}, and we include the Rh square lattice formed by ions at Wyckoff position $2b=(\tfrac12 \pm \tfrac14, \tfrac12 \mp \tfrac14,\tfrac12)$. The full Hamiltonian is given by
\begin{align}
    H = H_d + H_f + H_{fd},
    \label{eq:ham}
\end{align}
with $H_{d(f)}$ the $d(f)$-electron term and $H_{fd}$ the hybridization.
The $d$-electron Hamiltonian is
\begin{align}
    H_d = \sum_{\mathbf{k}} \mathbf{d}^\dagger_{\mathbf{k}}\, \mathcal{H}_d(\mathbf{k})\, \mathbf{d}_{\mathbf{k}},
\end{align}
with $\mathbf{d}_{\mathbf{k}}=(d_{\mathbf{k}A\uparrow},\,d_{\mathbf{k}A\downarrow},\,d_{\mathbf{k}B\uparrow},\,d_{\mathbf{k}B\downarrow})^{T}$, where $d_{\mathbf{k}\tau\sigma}$ is the annihilation operator for a $d$ electron located on the Rh $\tau=A,B$ sublattice with momentum $\mathbf{k}$ and spin $\sigma=\uparrow,\downarrow$. The Rh atoms form a square lattice rotated by $\pi/4$ with respect to the crystal axes. Including nearest- and next-nearest-neighbor spin-independent hopping terms gives the matrix elements
\begin{align}
    \mathcal{H}_d(\mathbf{k}) = {}&\left[\epsilon_d-2t_1(\cos k_x + \cos k_y)\right]\,\tau_0\sigma_0 \nonumber\\
    &- 4t_2\cos\tfrac{k_x}{2}\cos\tfrac{k_y}{2}\,\tau_x\sigma_0,
    \label{eq:Hd}
\end{align}
where $\tau_i$ and $\sigma_i$ are Pauli matrices in sublattice and spin space, respectively, $\epsilon_d$ is the onsite energy of the $d$ orbital, and $t_1$ ($t_2$) is the intra- (inter-)sublattice hopping arising from next-nearest- (nearest-)neighbor hopping.

The tetragonal crystal electric field (CEF) splits the $J=5/2$ states into three Kramers doublets given by \cite{Amorese2020,Christovam2024,Hafner2022}
\begin{align}
    \ket{\Gamma_6,\pm} &= \ket{\pm 1/2}  \label{eq:Gamma6} \\
    \ket{\Gamma_7^{(1)},\pm} &= \cos\theta \ket{\pm 5/2} - \sin\theta\ket{\mp 3/2}  \label{eq:Gamma71}\\
    \ket{\Gamma_7^{(2)},\pm} &= \sin\theta \ket{\pm 5/2} + \cos\theta\ket{\mp 3/2}, \label{eq:Gamma72}
\end{align}
in the $J=5/2$ basis $\ket{m_J}\equiv \ket{J=5/2, m_J}$, where the
angle $\theta$ fixes the mixing of the two $\Gamma_7$ doublets. Susceptibility measurements on CeRh$_2$As$_2$ give $\theta\simeq0.346\pi$ \cite{Hafner2022}. The $f$-electron Hamiltonian is entirely onsite, $H_f = \sum_{\mathbf{R},\tau} h_{\mathrm{loc}} (\mathbf{R}, \tau)$. In the CEF basis, $h_{\mathrm{loc}}$ is given by
\begin{align}
    h_{\mathrm{loc}} (\mathbf{R}, \tau) &= \sum_{\Gamma\sigma}\epsilon_\Gamma\, f^\dagger_{\mathbf{R}\tau\Gamma\sigma}f_{\mathbf{R}\tau\Gamma\sigma} \nonumber\\
    &+ \frac{1}{2} \sum_{\alpha\beta\gamma\delta} U_{\alpha\beta\gamma\delta}\, f^\dagger_{\mathbf{R}\tau\alpha}f^\dagger_{\mathbf{R}\tau\beta}f_{\mathbf{R}\tau\gamma}f_{\mathbf{R}\tau\delta},
    \label{eq:hloc}
\end{align}
where $\mathbf{R}$ runs over unit cells and the $f$-electron operators are defined analogously to the $d$-electron ones, with the addition of an orbital degree of freedom $\Gamma=\Gamma_6,\Gamma_7^{(1)},\Gamma_7^{(2)}$. Here $\tau=A,B$ labels the Ce sublattice, $\sigma=\pm$ is the Kramers pseudospin, $\epsilon_\Gamma$ is the CEF energy of doublet $\Gamma$, and $U_{\alpha\beta\gamma\delta}$ is the onsite Coulomb tensor acting locally on each Ce site $(\mathbf{R},\tau)$, where each of $\alpha,\beta,\gamma,\delta$ is a combined orbital and pseudospin index $(\Gamma,\sigma)$. The hybridization is given by
\begin{align}
    H_{fd} = \sum_{\mathbf{k},\Gamma}\left\{ \mathbf{d}^\dagger_{\mathbf{k}}\, V_\Gamma(\mathbf{k})\, \mathbf{f}_{\mathbf{k}\Gamma} + \mathrm{h.c.}\right\},
\end{align}
where $\mathbf{f}_{\mathbf{k}\Gamma}=(f_{\mathbf{k}\Gamma A+},\,f_{\mathbf{k}\Gamma A-},\,f_{\mathbf{k}\Gamma B+},\,f_{\mathbf{k}\Gamma B-})^{T}$. The form of the hybridization matrix depends on the choice of $d$ orbital and CEF doublet. Allowing for nearest-neighbor nonlocal $f$--$d$ hybridization \cite{Weber2008,Sourd2024}, for a $d_{x^2-y^2}$ conduction band, the hybridization matrix reads
\begin{align}
    V_\Gamma^\dagger(\mathbf{k}) = {}&2 v_1^\Gamma\!\left(i\cos\tfrac{k_x}{2}\,\tau_y \mp \cos\tfrac{k_y}{2}\,\tau_z\right)\sigma_0 \nonumber\\
    &+ 2 v_2^\Gamma\!\left(\sin\tfrac{k_x}{2}\,\tau_x\sigma_y \pm \sin\tfrac{k_y}{2}\,\tau_0\sigma_x\right),
    \label{eq:Vsk_x2y2}
\end{align}
where the upper (lower) sign denotes $\Gamma_7^{(1,2)}$ ($\Gamma_6$). The $\tau_i$ matrices connect the Rh and Ce sublattices, and $\sigma_i$ connect the $d$-spin and $f$-pseudospin. Thus, we refer to $v_{1}^\Gamma$ and $v_2^\Gamma$ as the spin-preserving and spin-flipping hybridizations, respectively, whose magnitudes depend on the chosen $\Gamma$ orbital. For $d_{xy}$, the same construction gives
\begin{align}
    V_\Gamma^\dagger(\mathbf{k}) = {}&2 v_1^\Gamma\!\left(\cos\tfrac{k_x}{2}\,\tau_y + i\cos\tfrac{k_y}{2}\,\tau_z\right)\sigma_z \nonumber\\
    &+ 2 v_2^\Gamma\!\left(\sin\tfrac{k_x}{2}\,\tau_x\sigma_x \mp \sin\tfrac{k_y}{2}\,\tau_0\sigma_y\right),
    \label{eq:Vsk}
\end{align}
where we drop the lower sign for the $v_1^\Gamma$ term since $v_1^{\Gamma_6}$ vanishes identically due to symmetry for $d_{xy}$. Because the $f$ and $d$ orbitals have opposite parity, the nonlocal $f$--$d$ hybridization for a centrosymmetric Ce site is odd in $\mathbf{k}$ \cite{Dzero2012}; the even-in-$\mathbf{k}$ $v_1^\Gamma$ is finite only due to the local breaking of inversion symmetry at the Ce site, which is captured within the Slater--Koster approach through the Ce--Rh bond angle. Although the local inversion-symmetry breaking in principle also permits onsite hybridization between the Ce $4f$ and $5d$ orbitals \cite{Gankhuyag2026,Matsumura2026}, we neglect this here, as first-principles calculations indicate that the low-energy conduction bands are predominantly of Rh $d$ character \cite{Ptok2021,Cavanagh2022,Ishizuka2023}.

Experimental measurements probing the CEF levels suggest a $\Gamma_7^{(1)}$ ground state with $\Gamma_6$ a closely competing excited state and $\Gamma_7^{(2)}$ well above the other two doublets \cite{Christovam2024,Hafner2022}. Accordingly, in the following we retain only the $\Gamma_7\equiv\Gamma_7^{(1)}$ and $\Gamma_6$ orbitals. Collecting the operators for the two orbitals into a single spinor $\mathbf{f}_{\mathbf{k}}=(\mathbf{f}_{\mathbf{k}A},\,\mathbf{f}_{\mathbf{k}B})^{T}$, with $\mathbf{f}_{\mathbf{k}\tau}=(f_{\mathbf{k}\tau\Gamma_7+},\,f_{\mathbf{k}\tau\Gamma_7-},\,f_{\mathbf{k}\tau\Gamma_6+},\,f_{\mathbf{k}\tau\Gamma_6-})^{T}$, the hybridization term is given by $H_{fd}=\sum_{\mathbf{k}}\{\mathbf{d}^\dagger_{\mathbf{k}}V(\mathbf{k})\mathbf{f}_{\mathbf{k}}+\mathrm{h.c.}\}$ with $[V(\mathbf{k})]_{(\tau\sigma),(\tau^\prime\Gamma\sigma^\prime)}=[V_\Gamma(\mathbf{k})]_{(\tau\sigma),(\tau^\prime\sigma^\prime)}$. DFT calculations suggest that the Rh $d_{xy}$ states give the dominant contribution at the Fermi level \cite{Cavanagh2022}, so we adopt Eq.~\eqref{eq:Vsk} in what follows.

\begin{table*}[t]
\caption{Here we tabulate the candidate magnetic orders that we consider during our numerical work. Each magnetic state consists of a primary dipole order and an induced quadrupole order. The ``/'' and ``$\pm$'' are used to display the two degenerate partner states which arise when considering two-dimensional irreps. Each of the magnetic states is shown schematically in Fig.~\ref{fig:states} and the full symmetry classification of all single-particle slave-boson channels is shown in Table~\ref{tab:irrep_grid}.}
\label{tab:states}
\begin{ruledtabular}
\begin{tabular}{lccccc}
& & \multicolumn{2}{c}{primary} & \multicolumn{2}{c}{induced}\\
\cline{3-4}\cline{5-6}
order & $\mathbf{Q}$ & irrep & dipole & irrep & quadrupole\\
\colrule
PM & $(0,0)$ & $A_{1g}$ & -- & -- & --\\
\colrule
FM$_z$ & $(0,0)$ & $A_{2g}$ & $J_z$ & -- & --\\
A-AFM$_z$ & $(0,0)$ & $A_{1u}$ & $J_z$ & -- & --\\
FM$_{x/y}$ & $(0,0)$ & $E_g^{(1)}\!\pm\!E_g^{(2)}$ & $J_x$ / $J_y$ & $B_{1g}$ & $O_{x^2-y^2}$\\
A-AFM$_{x/y}$ & $(0,0)$ & $E_u^{(1)}\!\pm\!E_u^{(2)}$ & $J_x$ / $J_y$ & $B_{1g}$ & $O_{x^2-y^2}$\\
FM$_{x\pm y}$ & $(0,0)$ & $E_g^{(1)}$ / $E_g^{(2)}$ & $J_x\!\pm\!J_y$ & $B_{2g}$ & $O_{xy}$\\
A-AFM$_{x\pm y}$ & $(0,0)$ & $E_u^{(1)}$ / $E_u^{(2)}$ & $J_x\!\pm\!J_y$ & $B_{2g}$ & $O_{xy}$\\
\colrule
orthorhombic $M_4$ & $(\pi,\pi)$ & $M_4^{(1)}$ / $M_4^{(2)}$ & $J_z$ & $B_{2g}$ & $O_{xy}$\\
orthorhombic $M_1$ & $(\pi,\pi)$ & $M_1^{(1)}$ / $M_1^{(2)}$ & $J_x\!\pm\!J_y$ & $B_{2g}$ & $O_{xy}$\\
orthorhombic $M_2$ & $(\pi,\pi)$ & $M_2^{(1)}$ / $M_2^{(2)}$ & $J_x\!\mp\!J_y$ & $B_{2g}$ & $O_{xy}$\\
tetragonal $M_4$ & $(\pi,\pi)$ & $M_4^{(1)}\!\pm\!M_4^{(2)}$ & $J_z, 0$ & $A_{2u}$ & $O_2^0$\\
tetragonal $M_1$ & $(\pi,\pi)$ & $M_1^{(1)}\!\pm\!M_1^{(2)}$ & $J_x, J_y$ & $B_{2u}$ & $O_{x^2-y^2}$\\
tetragonal $M_2$ & $(\pi,\pi)$ & $M_2^{(1)}\!\pm\!M_2^{(2)}$ & $J_x, J_y$ & $B_{2u}$ & $O_{x^2-y^2}$\\
\end{tabular}
\end{ruledtabular}
\end{table*}

\section{Rotationally invariant slave-boson mean-field theory}\label{sec:risb}

\subsection{General formalism}\label{sub:risb_general}

We treat the onsite repulsion of the $f$ electrons within the rotationally invariant slave-boson (RISB) formalism \cite{KotliarRuckenstein1986,Li1989,Lechermann2007,Piefke2018,Lanata2015,Lanata2017,Nourse2020,Riegler2020}. Following the standard procedure, the physical $f$-electron operator $\underline{f}$ is represented in an enlarged Hilbert space as
\begin{align}
\underline{f}_{i\alpha}=\hat{R}_{i,\alpha\beta}[\hat{\phi}]\,f_{i\beta},
\end{align}
where $\hat{R}_i$ is the renormalization matrix and $f_{i\beta}$ now represent the quasiparticle fermionic operators. In this work, we neglect all intersite correlations generated by the interaction such that $i=(\mathbf{R},\tau)$ enumerates the position of each Ce site and $\alpha,\beta=(\Gamma,\sigma)$ enumerate the four onsite fermionic degrees of freedom. The renormalization matrix is itself expressed in terms of the slave-boson operators $\hat{\phi}_{i,An}$ (see Ref.~\cite{Lechermann2007} for its explicit form), which carry two indices; a slave-boson operator is introduced for each pair $\{ \ket{A}, \ket{n} \}$ where $\ket{A}$ and $\ket{n}$ run over the $2^4=16$ states of the physical and quasiparticle local Fock spaces, respectively. Additionally, we must impose the following two constraints at every site $i$ to restrict our system to physical states
\begin{align}\label{eq:constraints}
    \sum_{A n} \hat{\phi}_{i,A n}^\dagger \hat{\phi}_{i,An} &= 1, \nonumber\\
    \sum_A \sum_{n n^\prime} \hat{\phi}_{i,A n^\prime}^\dagger \hat{\phi}_{i,An} \bra{n} f^\dagger_{\alpha} f_{\beta} \ket{n^\prime} &= f^\dagger_{i\alpha} f_{i\beta} ,
\end{align}
which are enforced by introducing a scalar Lagrange multiplier $\lambda_{0,i}$ for the norm and a Hermitian matrix $\Lambda_{i,\alpha\beta}$ for the density constraint. The total electron filling is also fixed by introducing a chemical potential $\mu$.

At the saddle point the bosons are replaced by their expectation values, collected into the amplitude matrix $[\Phi_\tau]_{An}=\langle\hat{\phi}_{i,An}\rangle$, which depends on position only through its sublattice index. The renormalization matrix is then a $c$-number, $\hat{R}_i[\hat{\phi}]\to R_\tau$, so that substituting for the physical $f$-electron operators in Eq.~\eqref{eq:ham} in terms of quasiparticle operators $\underline{f}_{\mathbf{R}\tau\alpha}=R_{\tau,\alpha\beta}f_{\mathbf{R}\tau\beta}$, and including contributions from enforcing the constraints Eq.~\eqref{eq:constraints}, gives a renormalized Hamiltonian which is bilinear in quasiparticle operators. The matrix elements of the full renormalized Hamiltonian are given by
\begin{align}\label{eq:Hrenorm}
    \mathcal{H}(\mathbf{k}) = \begin{pmatrix}
        \mathcal{H}_d(\mathbf{k}) & V(\mathbf{k}) R \\
        R^\dagger V^\dagger(\mathbf{k}) & \Lambda
    \end{pmatrix},
\end{align}
in the basis $(\mathbf{d}_{\mathbf{k}},\mathbf{f}_{\mathbf{k}})$ of Sec.~\ref{sec:model},
where $R \equiv \bigoplus_\tau R_\tau$ and $\Lambda \equiv \bigoplus_\tau \Lambda_\tau$ are block diagonal in the sublattice index. As seen below, the onsite $f$-electron Hamiltonian Eq.~\eqref{eq:hloc} is accounted for entirely through the slave-boson fields.

The solutions are then given by stationary points of the free-energy density
\begin{align}\label{eq:FE}
    \mathcal{F} = {}&-\frac{1}{N\beta}\sum_{\mathbf{k},j}\ln\!\left[1+e^{-\beta(\epsilon_{\mathbf{k},j}-\mu)}\right] + \mu n \nonumber\\
    &+ \sum_\tau\Big\{\Tr[\Phi_\tau\Phi_\tau^\dagger \Hloc] - \sum_{\alpha\beta}\Lambda_{\tau,\alpha\beta}\Tr[\Phi_\tau^\dagger\Phi_\tau \Fq{\alpha}^\dagger \Fq{\beta}] \nonumber\\
    &\qquad\quad + \lambda_{0,\tau}\big(\Tr[\Phi_\tau^\dagger\Phi_\tau] - 1\big)\Big\},
\end{align}
where $\epsilon_{\mathbf{k},j}$ are the eigenvalues of $\mathcal{H}(\mathbf{k})$, $\beta$ is the inverse temperature, $N$ the number of unit cells, and $n$ is the electron filling per unit cell. We also introduce the matrices $[\Hloc]_{AB}=\bra{A}h_{\mathrm{loc}}\ket{B}$ and $[\Fq{\alpha}]_{nm}=\bra{n}f_\alpha\ket{m}$, which represent the physical $f$-electron Hamiltonian and the quasiparticle annihilation operators in their respective onsite Fock bases. The stationary conditions are then given by
\begin{align}\label{eq:stationarity}
    \frac{\partial\mathcal{F}}{\partial\Phi^\dagger_\tau} = \frac{\partial\mathcal{F}}{\partial\lambda_{0,\tau}} = \frac{\partial\mathcal{F}}{\partial\Lambda_\tau} = \frac{\partial\mathcal{F}}{\partial\mu} = 0 .
\end{align}

The expectation values of local observables are computed similarly to how the local Hamiltonian, $\Hloc$, enters the free energy in Eq.~\eqref{eq:FE} \cite{Piefke2018}. For a given onsite operator $\hat{O}$, the corresponding expectation value on sublattice $\tau$ is given in terms of slave-boson amplitudes as
\begin{equation}\label{eq:O_loc}
  \langle\hat{O}\rangle_\tau=\Tr\!\left[\Phi_\tau \Phi_\tau^{\dagger} \fock{O} \right],
\end{equation}
where $[\fock{O}]_{AB}=\bra{A}\hat{O}\ket{B}$ is the matrix representation of the operator in the physical onsite Fock basis. In particular, for a one-body operator $\hat{O}=\sum_{\alpha\beta}O_{\alpha\beta}\underline{f}^\dagger_\alpha \underline{f}_\beta$, Eq.~\eqref{eq:O_loc} becomes
\begin{equation}\label{eq:rho_loc}
  \langle\hat{O}\rangle_\tau=\Tr\!\left[O\rho_\tau\right],
  \qquad
  \left[\rho_\tau\right]_{\alpha\beta}=\Tr\!\left[\Phi_\tau\Phi_\tau^{\dagger}\Cp{\beta}^{\dagger}\Cp{\alpha}\right],
\end{equation}
where $[\fock{C}_\alpha]_{AB}=\bra{A}\underline{f}_\alpha \ket{B}$ represent the physical electron operators, and $\rho_\tau$ is the local $f$-electron density matrix, which is gauge invariant by construction.

While the discussion above is restricted to $\mathbf{Q}=(0,0)$ states, it can easily be applied to $\mathbf{Q}=(\pi,\pi)$ states by considering the doubled magnetic unit cell and extending the site label to $(\eta,\tau)$, where $\eta=1,2$ corresponds to the additional magnetic sublattice degree of freedom, on which the Pauli matrices $\eta_i$ act. This is the approach we take when solving for the $\mathbf{Q}=(\pi,\pi)$ magnetic states discussed in the following sections.

Throughout our numerical work, we take the $U\to\infty$ limit, in which all configurations with more than one $f$ electron on a given site are projected out. Consequently, only the $N_f=0$ and $N_f=1$ sectors of the local Hilbert space remain, where $N_f$ denotes the onsite $f$-electron number. This approximation is physically motivated by the localized nature of the $f$ electrons, for which the Coulomb interaction is expected to be the dominant energy scale \cite{Christovam2024,Amorese2020}. The multiorbital nature of our model also makes this a pragmatic choice in terms of both model parameterization and computational cost. In particular, the multiorbital Coulomb interaction in Eq.~\eqref{eq:hloc} must encode the relative strengths of the active interaction channels, introducing additional parameters for each inequivalent channel. Moreover, the number of mean-field parameters entering Eqs.~\eqref{eq:FE} and \eqref{eq:stationarity} grows exponentially with the number of onsite degrees of freedom, reflecting the exponential growth of the local Fock space.

In this work, all saddle-point solutions are obtained on a lattice of $48\times48$ nonmagnetic unit cells. We define the free energy $\mathcal{F}$ and filling $n$ per nonmagnetic unit cell, so that states with different propagation vectors can be compared directly. We quote the free energy of an ordered state relative to the paramagnetic state at the same parameters and filling, $\dFE\equiv\mathcal{F}-\mathcal{F}_\mathrm{PM}$, and define $|\dFE|$ as its condensation energy.

\begin{figure*}[tp]
\includegraphics[width=\textwidth]{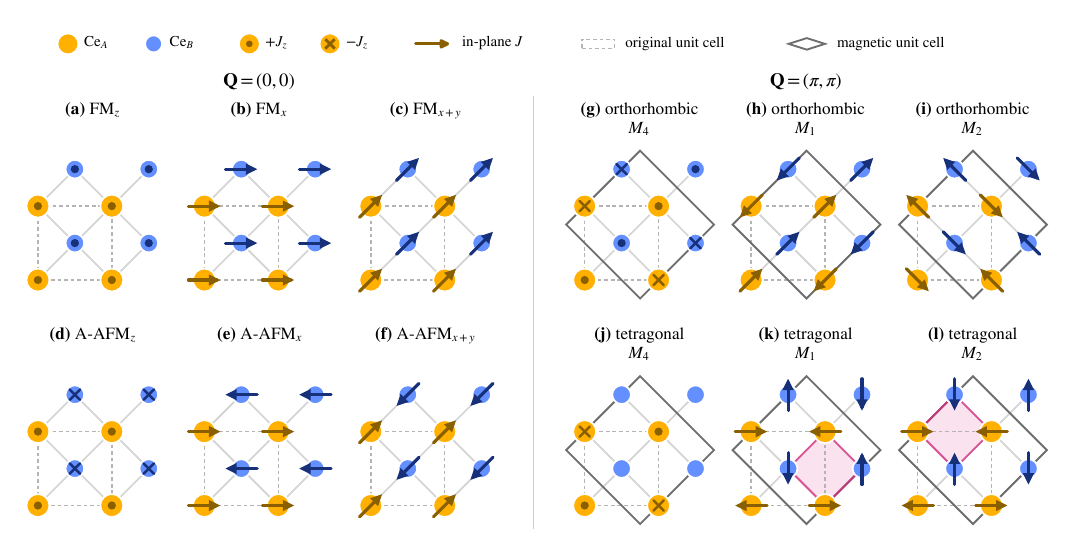}
\caption{The dipole moments on the Ce atoms for magnetic states in Table~\ref{tab:states}, as seen looking along the $c$ axis of the crystal. The shaded plaquettes for the tetragonal states highlight the vortex and all-in--all-out nature of the $M_1$ and $M_2$ states, respectively. }
\label{fig:states}
\end{figure*}

\subsection{Symmetry-restricted ans\"atze for magnetic states}\label{sub:ansatze}

The saddle-point equations derived from Eqs.~\eqref{eq:FE} and \eqref{eq:stationarity} do not uniquely specify a single solution. Indeed, solving these equations directly, one will find many solutions, including mixed-irrep solutions, depending on the choice of initial conditions \cite{Riegler2020}. As such, during a given solver run, we restrict our attention to a specific magnetic state which is identified by the spatial pattern of its magnetic dipole moment. Experimentally, NMR and NQR measurements point to $\mathbf{Q}=(0,0)$ A-type antiferromagnetic order with $c$-axis moments \cite{Kibune2022,Ogata2024,Ogata2026}, while inelastic neutron scattering finds magnetic fluctuations near $\mathbf{Q}=(\pi,\pi)$ \cite{TChen2024Expt}. We therefore provide a systematic study of $\mathbf{Q}=(0,0)$ and $\mathbf{Q}=(\pi,\pi)$ magnetic states whose dipole moments belong to a single irrep. The details of the states are provided in Table~\ref{tab:states}, and the corresponding dipole patterns are shown schematically in Fig.~\ref{fig:states}. Note that incommensurate order cannot be studied within our approach.

Once we choose the magnetic state we wish to study, the corresponding slave-boson ansatz can be constructed. Recalling that at $U\to\infty$ only the $N_f=0$ and $N_f=1$ sectors survive, on each site the amplitude matrix is block diagonal in $N_f$,
\begin{align}\label{eq:Phi_blocks}
    \Phi_{\eta\tau} = \begin{pmatrix} \Phib{0}_{\eta\tau} & 0\\ 0 & \Phib{1}_{\eta\tau}\end{pmatrix} ,
\end{align}
where $\Phib{0}_{\eta\tau}$ is a complex scalar and $\Phib{1}_{\eta\tau}$ is a $4\times4$ complex matrix. Since we neglect intersite correlations, we define the block-diagonal matrices $\Phib{n}\equiv\bigoplus_{\eta\tau}\Phib{n}_{\eta\tau}$ over the four sites of the magnetic unit cell. We then expand $\Phib{1}$ with complex coefficients in products $\eta_i\tau_j\alpha_k\sigma_l$, where $\alpha_k$ are Pauli matrices acting in the orbital space and the block-diagonal structure restricts the site matrices to $\eta_{0,z}$ and $\tau_{0,z}$. These products form single-particle slave-boson channels, which we label by irrep in Table~\ref{tab:irrep_grid}. An ansatz for a given magnetic state is then constructed by including only the channels of the relevant irreps,
\begin{align}\label{eq:Phi_ansatz}
    \Phib{1} = \underbrace{\Phib{1}_{\mathrm{PM}}}_{A_{1g}} + \underbrace{\Phib{1}_{\mathrm{prim}}}_{\text{dipole}} + \Phib{1}_{\mathrm{ind}} ,
\end{align}
where in all cases we include the paramagnetic $A_{1g}$ background, and $\Phib{1}_{\mathrm{prim}}$ and $\Phib{1}_{\mathrm{ind}}$ contain the channels of the primary and induced irreps in Table~\ref{tab:states}, respectively. The induced channels are included as the minimal extension which allows a particular ansatz to be closed under the saddle-point equations, and lead to induced quadrupolar moments in many of the magnetic states. The $N_f=0$ block is uniform across all sites, $\Phib{0}=\phib{0}_{A_{1g}}\eta_0\tau_0$, except when the $A_{2u}$ irrep is induced, where the staggered charge distribution requires an additional staggered channel, $\Phib{0}=\phib{0}_{A_{1g}}\eta_0\tau_0+\phib{0}_{A_{2u}}\eta_0\tau_z$.

Each magnetic state we consider here satisfies an antiunitary symmetry of the form $\mathcal{A} = g\,\mathcal{T}$, where $g$ is the space-group element which restores the dipole pattern after time reversal, e.g., $\eta_x \mathcal{T}$ symmetry for the $\mathbf{Q}=(\pi,\pi)$ states, with $g=\eta_x$ swapping the magnetic sublattices. The antiunitary symmetry must be satisfied by the renormalized quasiparticle Hamiltonian Eq.~\eqref{eq:Hrenorm}, which in turn places constraints on the $R$ matrices. We choose to enforce this symmetry directly on the slave-boson ansatz, from which the $R$ matrix inherits its form. This fixes the coefficient of each channel to be purely real or imaginary, giving a consistent parameterization of the magnetic state. Note that this parameterization is not unique due to the well-known gauge freedom in the slave-boson formalism \cite{Lechermann2007,Piefke2018}. The Lagrange multipliers $\Lambda\equiv\bigoplus_{\eta\tau}\Lambda_{\eta\tau}$ and $\lambda_{0}\equiv\bigoplus_{\eta\tau}\lambda_{0,\eta\tau}$ are restricted in the same way as $\Phib{1}$ and $\Phib{0}$, respectively, with the additional constraint of Hermiticity. We emphasize that the presence of an antiunitary symmetry does \emph{not} guarantee a Kramers degeneracy, which requires a $\mathcal{PT}$-like antiunitary symmetry such that arbitrary $\mathbf{k}$ is mapped to itself under the symmetry.

\begin{figure*}[t]
\includegraphics[width=\textwidth]{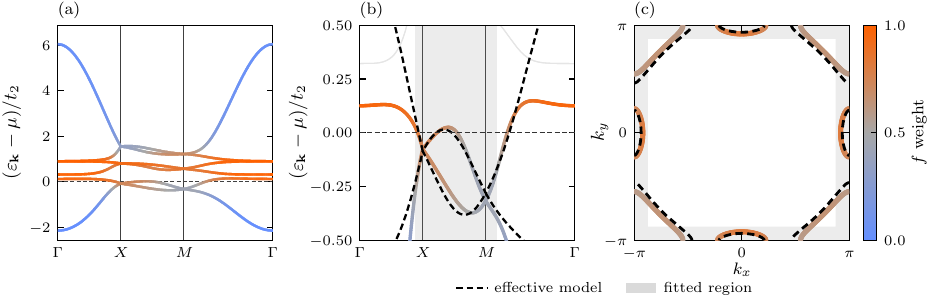}
\caption{The paramagnetic saddle point of the PAM and its low-energy effective $f$-electron description. (a) Band structure of the full renormalized PAM Eq.~\eqref{eq:Hrenorm}. (b) Low-lying bands of the PAM and the effective $f$-electron model Eq.~\eqref{eq:Heff_g7} fitted to them. (c) Fermi surfaces of the PAM and of the fitted effective model. Here, we let $t_2=1$ set the energy scale and solve self-consistently for the paramagnetic saddle point at fixed $t_1=-0.20t_2$, $\epsilon_{\Gamma_7}=-1.0t_2$, $\epsilon_{\Gamma_6}=-0.5t_2$, $v_1^{\Gamma_7}=0.30t_2$, $v_2^{\Gamma_7}=0.20t_2$, $v_1^{\Gamma_6}=0$ (symmetry enforced), $v_2^{\Gamma_6}=0.20t_2$, $n=2.30$, $\sqrt{N}=48$, $\beta=100/t_2$, and fit the effective model to the lowest PAM bands by least squares in the region $\max(|k_x|,|k_y|)\geq\pi-0.4$, which gives $\epsilon_f=-0.085t_2$, $t_{1}^f=-0.049t_2$, $|t_{2}^f|=0.185t_2$ and $|\alpha^f|=0.173t_2$.}
\label{fig:effective_fit}
\end{figure*}

As discussed above, the ans\"atze constrain the slave-boson fields on all sites of the unit cell. However, after choosing a particular ansatz and restricting the slave-boson amplitudes and Lagrange multipliers accordingly, we solve for the stationarity of the free-energy density in the original \emph{unrestricted} form, i.e., we ensure that the solutions we find are true solutions of Eq.~\eqref{eq:stationarity} where each site in the unit cell is treated independently. Although the stationarity condition is imposed in full, the restricted parameterization carries far fewer free parameters than the unrestricted problem, which makes the numerical search tractable \cite{Piefke2018}. Comparing the free energy obtained within each restricted ansatz, we can determine the dominant ordered state.

\section{Paramagnetic state}\label{sec:para}
\begin{figure*}[tp]
\includegraphics[width=\textwidth]{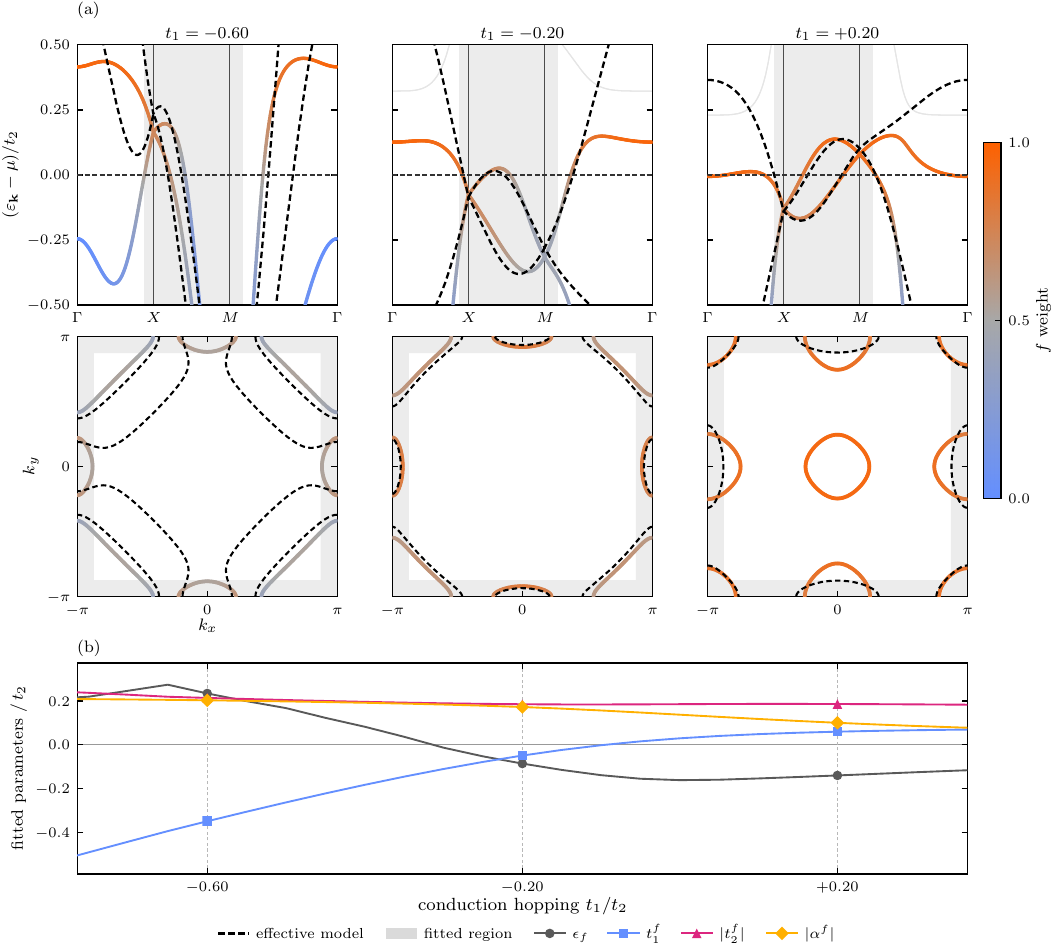}
\caption{Impact of the next-nearest-neighbor hopping of the conduction $d$ electrons, $t_1$, on the effective low-energy $f$-electron model Eq.~\eqref{eq:Heff_g7} fitted to the low-lying PAM bands as in Fig.~\ref{fig:effective_fit}. (a) Low-energy bands of the PAM and effective model fit (upper row), and the corresponding Fermi surfaces (lower row), at the varying values of $t_1$. (b) The fitted parameters of Eq.~\eqref{eq:Heff_g7} as we vary $t_1$. Here, we solve for the paramagnetic saddle point at fixed $\epsilon_{\Gamma_7}=-1.0t_2$, $\epsilon_{\Gamma_6}=-0.5t_2$, $v_1^{\Gamma_7}=0.30t_2$, $v_2^{\Gamma_7}=0.20t_2$, $v_1^{\Gamma_6}=0$, $v_2^{\Gamma_6}=0.20t_2$, $n=2.30$, $\sqrt{N}=48$, $\beta=100/t_2$, and fit the effective model in the region $\max(|k_x|,|k_y|)\geq\pi-0.4$.}
\label{fig:eff_panels_hop2}
\end{figure*}
We begin by discussing the model in the paramagnetic state. In this case, the renormalization of the quasiparticle Hamiltonian simplifies significantly as our ansatz includes only the $A_{1g}$ matrices and is uniform across all sites. The $\Phib{1}$ and $\Lambda$ matrices are given by
\begin{align}
    \Phib{1}_\mathrm{PM} = \eta_0\tau_0(\phib{1}_{A_{1g},1}\, \alpha_0 + \phib{1}_{A_{1g},2}\, \alpha_z)\sigma_0, \\
    \Lambda_\mathrm{PM} = \eta_0\tau_0(\lambda_{A_{1g},1}\,\alpha_0 + \lambda_{A_{1g},2}\, \alpha_z)\sigma_0,
\end{align}
where $\phib{1}_{A_{1g},a}$ and $\lambda_{A_{1g},a}$ ($a=1,2$) are real scalar parameters. Since the paramagnetic ansatz only contains orbital dependence, the renormalization independently scales the hybridization matrix for each orbital, i.e.,
\begin{align}
    V(\mathbf{k})\to V(\mathbf{k}) R_\mathrm{PM}= \big(V_{\Gamma_7}(\mathbf{k}) R_{\Gamma_7}, \ V_{\Gamma_6}(\mathbf{k}) R_{\Gamma_6}\big)
\end{align}
where $R_\Gamma$ are real scalars. An example band structure of a paramagnetic saddle point is shown in Fig.~\ref{fig:effective_fit}(a). We choose the noninteracting relative energy levels to be $\epsilon_{\Gamma_7} <\epsilon_{\Gamma_6}<\epsilon_d=0$ and solve self-consistently for the saddle point at a fixed electron filling $n=2.30$ per unit cell.

Most theoretical work on CeRh$_2$As$_2$ is based on a single-orbital effective $f$-electron model constructed from symmetry-allowed hoppings \cite{Khim2021,Mockli2021,Nogaki2021,Nogaki2022,Cavanagh2022,CLee2025RG}
\begin{align}\label{eq:Heff_g7}
    \mathcal{H}_\mathrm{eff}(\mathbf{k}) = {}& \epsilon_f\,\tau_0\sigma_0
      - 2t_{1}^f(\cos k_x + \cos k_y)\,\tau_0\sigma_0 \nonumber\\
    & - 4t_{2}^f \cos\tfrac{k_x}{2}\cos\tfrac{k_y}{2}\,\tau_x\sigma_0 \nonumber\\
    & + \alpha^f (\sin k_x\,\tau_z\sigma_y -\sin k_y\,\tau_z\sigma_x).
\end{align}
A model of this form can be obtained from our PAM by integrating out the conduction $d$ electrons. Neglecting the $\Gamma_6$ orbital and carrying out the Gaussian integral over the $d$ electrons, one obtains an effective $\Gamma_7$-only model
\begin{align}
  \Tilde{\mathcal{H}}_\text{eff}({\bf k}) = \epsilon_{\Gamma_7}\, \tau_0\sigma_0 + V_{\Gamma_7}^\dagger({\bf k})\,G_{d}({\bf k},\epsilon_{\Gamma_7})\,V_{\Gamma_7}({\bf k}),
\end{align}
where $G_{d}({\bf k},\epsilon_{\Gamma_7}) = \left[\epsilon_{\Gamma_7} - \mathcal{H}_d (\mathbf{k})\right]^{-1}$ is the Green function for the $d$ electrons, which we have analytically continued to the real axis and evaluated at the $\Gamma_7$ CEF energy level. Here we take the hybridization and the $\Gamma_7$ level to already include the slave-boson renormalization, which we do not show explicitly for clarity. Assuming the conduction band and the $f$-electron band are sufficiently separated, $|\Delta_{fd}|\gg 4(|t_1|+|t_2|)$ with $\Delta_{fd}\equiv\epsilon_{\Gamma_7} - \epsilon_d$, we may approximate the Green function as $G_{d}({\bf k},\epsilon_{\Gamma_7})\approx 1/\Delta_{fd}$. In this limit, the effective dispersion is controlled by $V_{\Gamma_7}^\dagger\,V_{\Gamma_7}$, which takes the same matrix form as the symmetry-based model in Eq.~\eqref{eq:Heff_g7}. Writing $v_i\equiv v_i^{\Gamma_7}$, we can express the parameters in Eq.~\eqref{eq:Heff_g7} in terms of the parameters of the PAM
\begin{align}\label{eq:eff_couplings}
    \epsilon_f &= \epsilon_{\Gamma_7} + \frac{4(v_1^2+v_2^2)}{\Delta_{fd}}, \nonumber\\
    t_1^f &= -\frac{v_1^2-v_2^2}{\Delta_{fd}}, \quad
    t_2^f = -\frac{2v_1^2}{\Delta_{fd}}, \quad
    \alpha^f = \frac{4 v_1 v_2}{\Delta_{fd}}.
\end{align}
Note that this correspondence is independent of whether the $d$ orbital is chosen to be $d_{xy}$ or $d_{x^2-y^2}$. Our approach provides a microscopic origin for the terms appearing in the effective model through $f$--$d$ hybridization. Both the spin-orbit coupling and intersublattice hopping of the effective model originate from the locally-broken inversion symmetry. Their ratio, $|\alpha^f/t_2^f|=2|v_2/v_1|$, is fixed by the relative strength of the two hybridization channels.

For the parameters we consider, however, the conduction and $f$ bands are not well separated, so that Eq.~\eqref{eq:eff_couplings} does not hold quantitatively. Nonetheless, we can use Eq.~\eqref{eq:Heff_g7} as an ansatz for the low-energy bands, treating its couplings as free parameters. The two lowest bands of the PAM, each doubly degenerate, have a dominant $\Gamma_7$ character near the Fermi energy. Fitting Eq.~\eqref{eq:Heff_g7} to these bands near the BZ edge reproduces both the low-energy band structure and the Fermi surface well, as shown in Figs.~\ref{fig:effective_fit}(b) and (c). Both the PAM and the fitted model exhibit Van Hove singularities, which have been observed near the edge of the BZ experimentally \cite{Ishizuka2023,XChen2024PRX,BChen2024Expt,Wu2024Expt}, and have been studied in the effective model \cite{CLee2025RG}. Our fit in Fig.~\ref{fig:effective_fit} gives $|t_2^f| \sim |\alpha^f| > |t_1^f|$; however, the low-energy band structure is sensitive to the dispersion of the conduction electrons, and the fit breaks down away from the BZ edge.s

The effect of tuning the next-nearest-neighbor hopping of the $d$ electrons, $t_1$ in Eq.~\eqref{eq:Hd}, is shown in Fig.~\ref{fig:eff_panels_hop2}. We see that the strengths of the intrasublattice hopping and the spin-orbit coupling vary as we tune $t_1$, whereas $|t_2^f|$ is stable.
At $t_1=+0.20t_2$, we find an additional $f$-electron-like Fermi surface near the $\Gamma$ point, but the bands near the BZ edge remain well described by the fitted model. For $t_1\lesssim-0.3t_2$, by contrast, the conduction band crosses the Fermi level near the BZ edge. The low-energy bands then acquire substantial $d$ character, and at $t_1=-0.60t_2$ the fitted model no longer reproduces the Fermi surface. The breakdown of the fit highlights that Eq.~\eqref{eq:Heff_g7} neglects the interplay between the conduction and $f$ electrons, which our PAM captures.

\begin{figure*}[tp]
\includegraphics[width=\textwidth]{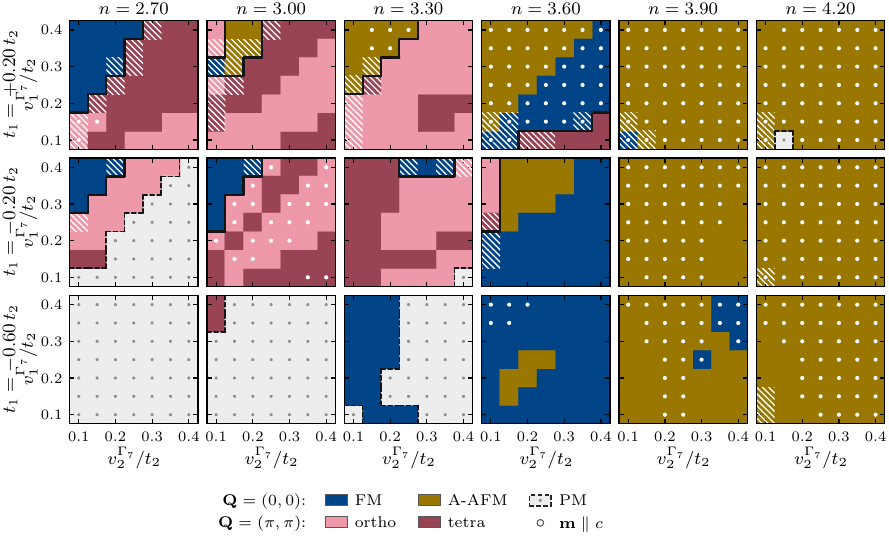}
\caption{Ground-state magnetic order across the hybridization plane $(v_2^{\Gamma_7}, v_1^{\Gamma_7})$, at varying electron filling $n$ (columns) and next-nearest-neighbor hopping of the conduction $d$ electrons, $t_1$ (rows). Each cell is colored by the lowest free-energy state among the candidate orders of Table~\ref{tab:states}: the $\mathbf{Q}=(0,0)$ orders are grouped by dipole pattern into FM and A-AFM, and the $\mathbf{Q}=(\pi,\pi)$ orders into orthorhombic and tetragonal (see Fig.~\ref{fig:states}). A white dot marks the cells whose ground state has its moments along the $c$ axis. Hatching marks the cells in which the lowest $\mathbf{Q}=(0,0)$ and $\mathbf{Q}=(\pi,\pi)$ states lie within $10^{-3}t_2$ of one another. Solid black lines separate regions of different $\mathbf{Q}$, while dashed lines bound the cells marked PM, in which no candidate order lies below the paramagnet. Here, we fix $\epsilon_{\Gamma_7}=-1.0t_2$, $\epsilon_{\Gamma_6}=-0.5t_2$, $v_1^{\Gamma_6}=0$, $v_2^{\Gamma_6}=0.20t_2$, $\sqrt{N}=48$, $\beta=100/t_2$, and solve for the saddle point of each candidate magnetic state.}
\label{fig:catalog_state_map}
\end{figure*}

\begin{figure*}[tp]
\includegraphics[width=\textwidth]{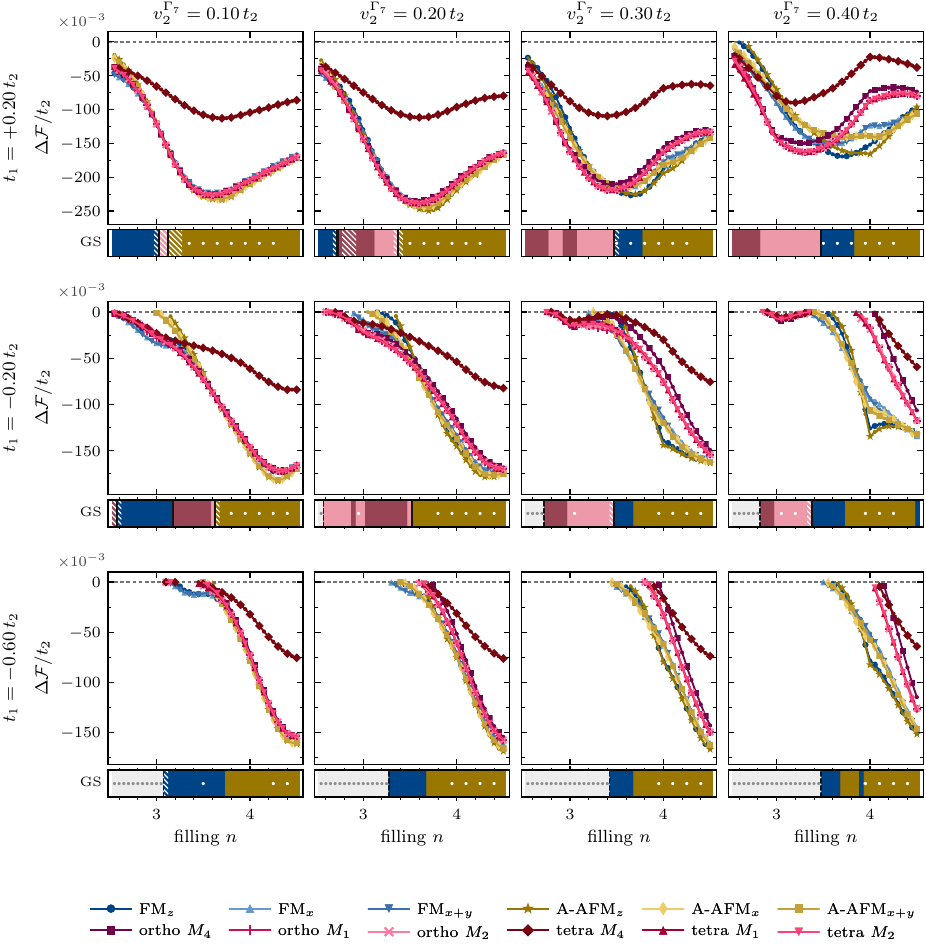}
\caption{Free energy of each candidate magnetic order of Table~\ref{tab:states} relative to the paramagnet, $\dFE=\mathcal{F}-\mathcal{F}_\mathrm{PM}$ in units of $t_2$ and negative where the order is stable, against electron filling $n$, at varying hybridization $v_2^{\Gamma_7}$ (columns) and next-nearest-neighbor hopping of the conduction $d$ electrons, $t_1$ (rows). The ground-state strip beneath each panel marks the lowest group at each filling, and uses the same colors, white dots and hatching as the phase diagram, Fig.~\ref{fig:catalog_state_map}. Here, we fix $\epsilon_{\Gamma_7}=-1.0t_2$, $\epsilon_{\Gamma_6}=-0.5t_2$, $v_1^{\Gamma_7}=0.30t_2$, $v_1^{\Gamma_6}=0$, $v_2^{\Gamma_6}=0.20t_2$, $\sqrt{N}=48$, $\beta=100/t_2$, and solve for the saddle point of each candidate magnetic state.}
\label{fig:catalog_dFE}
\end{figure*}

\section{Magnetic ordering}\label{sec:magnetic}

\subsection{Phase diagram}\label{sub:phase_diagram}
We now study the magnetic phase diagram of our model \cite{Dorin1992,Moeller1993,Doradzinski1998,Riegler2020}, considering all of the candidate magnetic states listed in Table~\ref{tab:states}. Following the method outlined in Sec.~\ref{sub:ansatze}, we solve for the saddle point within each restricted ansatz and compare the resulting free energies to determine the ground state. The main result of this section is presented in Fig.~\ref{fig:catalog_state_map}. In each panel we fix $t_1$ and $n$, and characterize the ground state in the $(v_2^{\Gamma_7}, v_1^{\Gamma_7})$-plane. We group the $\mathbf{Q}=(0,0)$ states as FM and A-AFM, and the $\mathbf{Q}=(\pi,\pi)$ states as orthorhombic or tetragonal, and mark the cells in which the ground-state moments lie along the $c$ axis. Solid black lines mark the phase boundaries between states of different $\mathbf{Q}$, and we hatch those cells in which states of different $\mathbf{Q}$ compete closely. In regions of parameter space where we cannot resolve the propagation vector of the ground state, our ans\"atze are likely to be insufficient; the system may, for example, prefer to order at an incommensurate wave vector \cite{CLee2026,Ogata2026,Galeski2026}.

The phase diagram separates into two regimes, set primarily by the filling: $\mathbf{Q}=(\pi,\pi)$ order at low filling and $\mathbf{Q}=(0,0)$ order at high filling. For $n\lesssim3.3$ and $t_1\ge-0.20t_2$, the ground state is $\mathbf{Q}=(\pi,\pi)$ order, except for $v_1^{\Gamma_7}\gtrsim1.5\,v_2^{\Gamma_7}$, where $\mathbf{Q}=(0,0)$ order takes over. The orthorhombic and tetragonal states compete closely throughout this regime (see Fig.~\ref{fig:catalog_dFE}). In most of this region the moments are directed in-plane, although at $n=3.00$ and $t_1=-0.20t_2$ we find out-of-plane order, predominantly the orthorhombic $M_4$ state. By $n=3.60$, $\mathbf{Q}=(0,0)$ order covers most of the plane; $\mathbf{Q}=(\pi,\pi)$ order survives only at small $v_1^{\Gamma_7}$ for $t_1=+0.20t_2$ and at small $v_2^{\Gamma_7}$ for $t_1=-0.20t_2$, and is absent for $n\gtrsim3.9$. Near the crossover between the two regimes, $n\approx3.3$--$3.6$, FM order appears. For $n\gtrsim3.9$, A-AFM order with predominantly out-of-plane moments (i.e. A-AFM$_z$) takes over at all three values of $t_1$.

Figure~\ref{fig:catalog_dFE} shows the free energy relative to the paramagnet, $\dFE$, of each magnetic state as a function of filling, for a range of $t_1$ and $v_2^{\Gamma_7}$. The conduction-electron hopping $t_1$, which reshapes the low-energy band structure (Fig.~\ref{fig:eff_panels_hop2}), has a major effect: decreasing $t_1$ from positive to negative values weakens the magnetic order, most strongly at low filling. The $\mathbf{Q}=(\pi,\pi)$ regime therefore shrinks with decreasing $t_1$ and has almost disappeared at $t_1=-0.60t_2$, where the system remains largely paramagnetic up to $n\approx3.3$ and then orders at $\mathbf{Q}=(0,0)$. In contrast, the A-AFM order near $n=4$ survives for all three values of $t_1$: decreasing $t_1$ from $+0.20t_2$ to $-0.60t_2$ reduces its condensation energy by about half. The spin-flip hybridization $v_2^{\Gamma_7}$ tends to destabilize the magnetic order, pushing its onset to higher fillings.  At $t_1=-0.20t_2$, the onset of the $\mathbf{Q}=(0,0)$ states shifts furthest, consistent with their suppression by $v_2^{\Gamma_7}$ in the phase diagram, Fig.~\ref{fig:catalog_state_map}, whereas at $t_1=-0.60t_2$ it is the onset of the $\mathbf{Q}=(\pi,\pi)$ states that shifts furthest. 

States with the same propagation vector compete closely. At small $v_2^{\Gamma_7}$ they are poorly resolved in energy, and they become well separated as $v_2^{\Gamma_7}$ increases. One mechanism for this is the effective spin-orbit coupling of the low-energy theory, which, from Eq.~\eqref{eq:eff_couplings}, vanishes if either hybridization channel is absent.
The exception is the tetragonal $M_4$ state, which is well resolved from the other $\mathbf{Q}=(\pi,\pi)$ states even at small $v_2^{\Gamma_7}$, as it is the only state in which the moment orders on just half of the Ce sites. Among the $\mathbf{Q}=(0,0)$ states, FM order sets in at lower filling than A-AFM and remains lower in energy up to $n\approx3.7$, which gives the FM ground state near the crossover in Fig.~\ref{fig:catalog_state_map}. At higher filling, FM and A-AFM are nearly degenerate, increasingly so as $t_1$ becomes more negative. This suggests that the two Ce sublattices order almost independently, coupled only by a weak antiferromagnetic exchange. 

Finally, the free energy shows distinct features at commensurate fillings. At $t_1=-0.20t_2$ and larger $v_2^{\Gamma_7}$, the $\mathbf{Q}=(\pi,\pi)$ states form a shallow dome of stability near $n=3$, where the orthorhombic $M_4$, orthorhombic $M_1$ and tetragonal $M_2$ states are filling-enforced magnetic Dirac semimetals \cite{Watanabe2016,Tang2016,Young2017,Watanabe2018} and the tetragonal $M_1$ state is a small-gap insulator. The $\mathbf{Q}=(0,0)$ curves show a similar feature at $n=4$, clearest at $v_2^{\Gamma_7}=0.40t_2$, where the FM$_z$ and A-AFM$_z$ states become insulating. In both cases, moving quasiparticle weight away from the Fermi level appears to stabilize the magnetic order.

\begin{figure}[tp]
\includegraphics[width=\columnwidth]{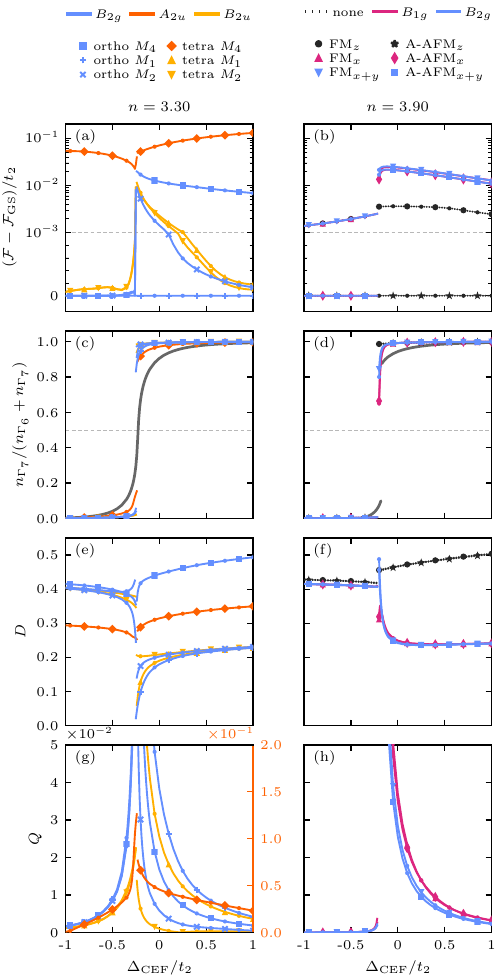}
\caption{Dependence of the magnetic states on the CEF splitting $\DCEF=\epsilon_{\Gamma_6}-\epsilon_{\Gamma_7}$, for the $\mathbf{Q}=(\pi,\pi)$ states at $n=3.30$ (left) and the $\mathbf{Q}=(0,0)$ states at $n=3.90$ (right). (a,b) Free energy relative to the ground state, plotted on a linear scale below the dashed line at $10^{-3}t_2$ and a logarithmic scale above it. (c,d) $\Gamma_7$ occupancy fraction $n_{\Gamma_7}/(n_{\Gamma_6}+n_{\Gamma_7})$, with the paramagnetic state in gray. (e,f) Dipole magnitude $D$. (g,h) Quadrupole magnitude $Q$, with the tetragonal $M_4$ state plotted on the right-hand axis in (g). Line colors indicate the induced quadrupole of each state (Table~\ref{tab:states}), and dotted lines mark states without one. Here, we fix $t_1=+0.20t_2$, $(\epsilon_{\Gamma_6}+\epsilon_{\Gamma_7})/2=-0.75t_2$, $v_1^{\Gamma_7}=0.30t_2$, $v_2^{\Gamma_7}=0.30t_2$, $v_1^{\Gamma_6}=0$, $v_2^{\Gamma_6}=0.20t_2$, $\sqrt{N}=48$, $\beta=100/t_2$.}
\label{fig:quad_delta}
\end{figure}

\subsection{Induced quadrupole order}\label{sec:quadrupole}

We now turn our attention to the induced quadrupole order in our model. We wish to determine the significance of the quadrupole moment relative to the primary dipole order. To this end, we construct normalized magnitudes for the dipole and quadrupole orders. For a given onsite single-particle operator $O$, consider its traceless projection onto the low-energy $\Gamma_7$--$\Gamma_6$ subspace
\begin{align}\label{eq:Oeff}
      O_\mathrm{proj} &=POP-\tfrac{1}{4}\Tr\!\left[POP\right]I_{4},
\end{align}
where $P$ is the projection operator
\begin{align}\label{eq:projector}
  P&=\sum_{\Gamma=\Gamma_7,\Gamma_6}\ \sum_{\sigma=\pm}\ket{\Gamma,\sigma}\!\bra{\Gamma,\sigma}.
\end{align}
Following Eq.~\eqref{eq:rho_loc}, the normalized expectation values in the magnetic unit cell are given by
\begin{equation}\label{eq:expectation}
  \langle O\rangle_{ij}=\frac{1}{4\,\Tr\!\left[O_\mathrm{proj}^\dagger O_\mathrm{proj}\right]^{1/2}}\,\Tr\!\left[\left(\eta_i\tau_j\otimes O_\mathrm{proj}\right)\rho\right],
\end{equation}
where $i,j\in\{0,z\}$ project onto the four-site sign patterns of Table~\ref{tab:irrep_grid}, and $\rho\equiv\bigoplus_{\eta\tau}\rho_{\eta\tau}$ is the site-resolved $f$-electron density matrix of the magnetic unit cell. Collecting the dipole and quadrupole operators as $\mathcal{D}=\{J_x,\,J_y,\,J_z\}$ and $\mathcal{Q}=\{O_2^0,\,O_{x^2-y^2},\,O_{xy},\,O_{yz},\,O_{zx}\}$, respectively, we define the corresponding magnitudes as
\begin{align}\label{eq:QD_def}
  D&= \sqrt{\sum_{O\in\mathcal{D}}\ \sum_{ij}\langle O\rangle_{ij}^{2}},\\
  Q&=\sqrt{\sum_{O\in\mathcal{Q}}\ \sum_{ij}\langle O\rangle_{ij}^{2}-\langle O_2^0\rangle_{00}^{2}},
\end{align}
where we subtract the trivial $A_{1g}$ contribution in the quadrupole magnitude, which is finite already in the paramagnetic state. Note that $O_2^0$ is purely intraorbital, whereas $O_{x^2-y^2}$, $O_{xy}$, $O_{yz}$ and $O_{zx}$ are purely $\Gamma_7$--$\Gamma_6$ interorbital, and therefore can only have finite expectation values when both orbitals are occupied. Explicit definitions of the dipole and quadrupole operators in the $J=5/2$ manifold are given in Ref.~\cite{Schmidt2024}.

At $t_1=+0.20t_2$, we consider one filling from each regime of the phase diagram: $n=3.30$, where the ground state is $\mathbf{Q}=(\pi,\pi)$ order, and $n=3.90$, where it is $\mathbf{Q}=(0,0)$ order. In Fig.~\ref{fig:quad_delta}, we show the free energy of each state relative to the ground state, the $\Gamma_7$ occupancy fraction, and the dipole and quadrupole magnitudes as we tune the CEF splitting $\DCEF\equiv\epsilon_{\Gamma_6}-\epsilon_{\Gamma_7}$. Most of these quantities show a discontinuous change as we tune from a $\Gamma_6$- to $\Gamma_7$-CEF ground state with increasing $\DCEF$. As seen in the occupancy fraction in Figs.~\ref{fig:quad_delta}(c) and (d), the critical $\DCEF$ at which the two orbitals are equally populated is shifted from $\DCEF=0$, even in the paramagnetic state. The shift arises from the unequal renormalization of the CEF level energies by hybridization with the conduction electrons, as illustrated by Eq.~\eqref{eq:eff_couplings} for the low-energy effective theory.

The free energy relative to the ground state, shown in Figs.~\ref{fig:quad_delta}(a) and (b), exhibits a pronounced asymmetry between the $\Gamma_6$- and $\Gamma_7$-dominated regimes, with essentially vanishing magnetic anisotropy in the former. This can also be understood from the low-energy effective theory: since the spin-preserving hybridization channel vanishes for the $\Gamma_6$ orbital, the effective spin-orbit coupling generated from $\Gamma_6$--$d$ hybridization vanishes. Spin-orbit coupling in the low-energy theory for the $\Gamma_6$ electrons can arise only through $d$-electron-mediated $\Gamma_7$--$\Gamma_6$ coupling, and is therefore expected to be much weaker than that observed for the $\Gamma_7$ electrons.

As shown in Figs.~\ref{fig:quad_delta}(e) and (f), near the orbital crossover the dipole moments of several $\mathbf{Q}=(\pi,\pi)$ states at $n=3.30$ are suppressed, whereas those of the FM$_z$ and A-AFM$_z$ states at $n=3.90$ remain almost constant. Reducing the separation of the $\Gamma_7$ and $\Gamma_6$ levels therefore weakens the $\mathbf{Q}=(\pi,\pi)$ order. At both fillings, the magnetic states switch from $\Gamma_6$- to $\Gamma_7$-dominated through a first-order transition.

As we approach the transition, the quadrupole moments, shown in Figs.~\ref{fig:quad_delta}(g) and (h), are strongly enhanced by the growing occupancy of the subdominant orbital that all quadrupoles, other than $A_{2u}$, require to be finite. Of particular interest is the orthorhombic $M_1$ state, which shows the largest quadrupole moment as we approach the crossover from the $\Gamma_7$-dominated side. On this side it is also the ground state at $n=3.30$, and close to the crossover its dipole moment is so strongly suppressed that it falls below the quadrupole moment. This behavior exemplifies that the quadrupole magnitude does \emph{not} follow the naive $Q\propto D^2$ one might expect for an induced order; however, in no case do we find that the quadrupolar moment is present without a dipole moment. At $n=3.90$, the ground state on the $\Gamma_7$-dominated side is A-AFM$_z$, which carries no induced quadrupole. The in-plane $\mathbf{Q}=(0,0)$ states acquire large quadrupoles near the crossover, but remain well above the ground state in energy. While the quadrupole magnitude can become significant for the ordered states, it requires $\DCEF$ to be fine-tuned near the critical point of the $\Gamma_7$--$\Gamma_6$ crossover, which suggests it may not play a critical role in shaping the zero-field phase diagram.

\section{Discussion}\label{sec:discussion}

We have studied a multiorbital periodic Anderson model for CeRh$_2$As$_2$ that retains both the $\Gamma_7\oplus\Gamma_6$ quasi-quartet and the Rh $d$ conduction states, treating the onsite correlations within the RISB approach. In the paramagnetic state, the minimal model of Eq.~\eqref{eq:Heff_g7} emerges as the low-energy description when the conduction band is well separated from the $f$ levels, providing a microscopic origin for its spin-orbit coupling, which arises from the interplay of $f$--$d$ hybridization and local inversion-symmetry breaking.

Using symmetry-restricted ans\"atze for the candidate $\mathbf{Q}=(0,0)$ and $\mathbf{Q}=(\pi,\pi)$ states, we find that the ordering wave vector is set primarily by the filling: in-plane $\mathbf{Q}=(\pi,\pi)$ order is favored at low filling, and A-AFM$_z$ order, with FM$_z$ close in energy, for $n\gtrsim3.9$. Our low-filling $\mathbf{Q}=(\pi,\pi)$ order is consistent with the magnetic fluctuations near this wave vector observed by inelastic neutron scattering \cite{TChen2024Expt}. However, these fluctuations are insensitive to the onset of phase I and no static order at this wave vector has been observed. In contrast, the A-AFM$_z$ state found at high fillings matches the order inferred from NQR measurements \cite{Kibune2022}, which has since been supported by subsequent NMR and NQR studies \cite{Ogata2024,Ogata2026}. The near-degeneracy of the A-AFM and FM states further indicates that the order consists of ferromagnetic Ce layers which are only weakly coupled to each other. The robust preference for $c$-axis polarization is nevertheless surprising, as the magnetic susceptibility in the paramagnetic state is larger in the $ab$ plane than along the $c$ axis \cite{Khim2021}. Although the quasi-quartet structure of the CEF levels has motivated proposals that phase I carries quadrupolar character \cite{Hafner2022,Schmidt2024,Thalmeier2025}, we find that the quadrupole moments become large only near the $\Gamma_7$--$\Gamma_6$ crossover.

In the interests of computational tractability, we have only considered commensurate magnetic orders. However, the most recent NQR measurements find the A-type order superposed with a two-dimensional incommensurate modulation of in-plane moments \cite{Ogata2026}; incommensurate order in phase I has also been inferred by $\mu$SR and ultrasound experiments~\cite{Galeski2026,Khim2025uSR}. Incommensurate order may also arise from the near-degeneracy of the A-AFM and FM states. Specifically, the FM$_z$ ($A_{2g}$) and A-AFM$_z$ ($A_{1u}$) order parameters admit a Lifshitz invariant $M_{A,z}\partial_{z}M_{B,z}-M_{B,z}\partial_{z}M_{A,z}$, where $M_{\tau,z}$ is the $c$-axis moment on the Ce $\tau=A,B$ sublattice. This term is part of $\mathbf{M}_{A}\cdot\partial_{z}\mathbf{M}_{B}-\mathbf{M}_{B}\cdot\partial_{z}\mathbf{M}_{A}$, which is allowed by the non-relativistic spin space group \cite{Xiao2024}, and so does not rely on spin-orbit coupling to be sizeable. When FM$_z$ and A-AFM$_z$ fluctuations are comparable near the transition, this term can drive an incommensurate modulation of the order along the $c$ axis \cite{CLee2026}. If the near-degeneracy we find at zero temperature reflects such competing fluctuations, our results point toward incommensurate order modulated along the $c$ axis.

Finally, we briefly compare our results to previous theoretical treatments. At low filling, where Eq.~\eqref{eq:Heff_g7} describes the low-energy bands well, the in-plane $\mathbf{Q}=(\pi,\pi)$ order we find agrees with weak-coupling studies of the minimal model, which find dominant in-plane $\mathbf{Q}=(\pi,\pi)$ fluctuations \cite{Nogaki2022}. A periodic Anderson model has also previously been applied to CeRh$_2$As$_2$ within a weak-coupling approach, finding dominant A-AFM$_z$ and closely-competing subdominant FM$_z$ fluctuations \cite{Nogaki2026}. It would be interesting to explore whether our similar conclusions share a common mechanism. In contrast, DFT+DMFT calculations find dominant out-of-plane $\mathbf{Q}=(\pi,\pi)$ fluctuations \cite{Numa2026}, whereas such order is rarely the ground state in our model. This calculation, however, treats the $4f$ electrons as localized, which limits its applicability to our model.

In summary, our work provides a framework for studying the ordered states of CeRh$_2$As$_2$ that retains both its multiorbital and its heavy-fermion character. In particular, our approach allows us to explore the interplay of magnetic and quadrupole orders within a strongly interacting itinerant electron picture.

\acknowledgments
This work was supported by the Marsden Fund Council from Government funding, managed by Royal Society Te Ap\={a}rangi, Contract No. UOO2222.

\bibliographystyle{apsrev4-2}
\bibliography{ref}
\vfill\eject\clearpage

\appendix

\section{Symmetry analysis of the slave-boson matrix}\label{app:ansatze}
\begin{table*}[t]
\caption{Symmetry labeling of single-particle slave-boson channels. Every channel is given by the product of an onsite $\alpha_k\sigma_l$ matrix with a four-site sign pattern $\eta_0\tau_{0,z}$ or $\eta_z\tau_{0,z}$, which give rise to $\mathbf{Q}=(0,0)$ or $\mathbf{Q}=(\pi,\pi)$ ordering, respectively. Here $\sigma_\pm\equiv\sigma_x\pm\sigma_y$. Superscripts on the two-dimensional irreps ($E_g$, $E_u$, and all $M$-point irreps) distinguish the two components of each irrep. We also include each channel's parity under time-reversal $\mathcal{T} = \eta_0\tau_0\alpha_0 (i\sigma_y) K$, where $K$ denotes complex conjugation, together with the onsite multipole (here we consider only dipole and quadrupole moments) onto which the channel has a finite projection. }
\label{tab:irrep_grid}
\begin{ruledtabular}
\begin{tabular}{llccccc}
 & & & \multicolumn{2}{c}{$\mathbf{Q}=(0,0)$} & \multicolumn{2}{c}{$\mathbf{Q}=(\pi,\pi)$}\\
channel & onsite multipole & $\mathcal{T}$ & $\eta_0\tau_0$ & $\eta_0\tau_z$ & $\eta_z\tau_0$ & $\eta_z\tau_z$\\
\colrule
$\alpha_{0}\sigma_0$ & $O_2^0$ & $+$ & $A_{1g}$ & $A_{2u}$ & $M_3^{(1)}$ & $M_3^{(2)}$\\
$\alpha_{z}\sigma_0$ & $O_2^0$ & $+$ & $A_{1g}$ & $A_{2u}$ & $M_3^{(1)}$ & $M_3^{(2)}$\\
$\alpha_{x}\sigma_0$ & $O_{x^2-y^2}$ & $+$ & $B_{1g}$ & $B_{2u}$ & $M_4^{(1)}$ & $M_4^{(2)}$\\
$\alpha_{y}\sigma_0$ & -- & $-$ & $B_{1g}$ & $B_{2u}$ & $M_4^{(1)}$ & $M_4^{(2)}$\\
\colrule
$\alpha_{0}\sigma_z$ & $J_z$ & $-$ & $A_{2g}$ & $A_{1u}$ & $M_4^{(1)}$ & $M_4^{(2)}$\\
$\alpha_{z}\sigma_z$ & $J_z$ & $-$ & $A_{2g}$ & $A_{1u}$ & $M_4^{(1)}$ & $M_4^{(2)}$\\
$\alpha_{x}\sigma_z$ & -- & $-$ & $B_{2g}$ & $B_{1u}$ & $M_3^{(1)}$ & $M_3^{(2)}$\\
$\alpha_{y}\sigma_z$ & $O_{xy}$ & $+$ & $B_{2g}$ & $B_{1u}$ & $M_3^{(1)}$ & $M_3^{(2)}$\\
\colrule
$\alpha_{0}\sigma_+$ & $J_x+J_y$ & $-$ & $E_{g}^{(1)}$ & $E_{u}^{(2)}$ & $M_1^{(1)}$ & $M_2^{(2)}$\\
$\alpha_{z}\sigma_+$ & $J_x+J_y$ & $-$ & $E_{g}^{(1)}$ & $E_{u}^{(2)}$ & $M_1^{(1)}$ & $M_2^{(2)}$\\
$\alpha_{x}\sigma_+$ & $J_x-J_y$ & $-$ & $E_{g}^{(2)}$ & $E_{u}^{(1)}$ & $M_2^{(1)}$ & $M_1^{(2)}$\\
$\alpha_{y}\sigma_+$ & $O_{yz}+O_{zx}$ & $+$ & $E_{g}^{(2)}$ & $E_{u}^{(1)}$ & $M_2^{(1)}$ & $M_1^{(2)}$\\
\colrule
$\alpha_{0}\sigma_-$ & $J_x-J_y$ & $-$ & $E_{g}^{(2)}$ & $E_{u}^{(1)}$ & $M_2^{(1)}$ & $M_1^{(2)}$\\
$\alpha_{z}\sigma_-$ & $J_x-J_y$ & $-$ & $E_{g}^{(2)}$ & $E_{u}^{(1)}$ & $M_2^{(1)}$ & $M_1^{(2)}$\\
$\alpha_{x}\sigma_-$ & $J_x+J_y$ & $-$ & $E_{g}^{(1)}$ & $E_{u}^{(2)}$ & $M_1^{(1)}$ & $M_2^{(2)}$\\
$\alpha_{y}\sigma_-$ & $O_{yz}-O_{zx}$ & $+$ & $E_{g}^{(1)}$ & $E_{u}^{(2)}$ & $M_1^{(1)}$ & $M_2^{(2)}$\\
\end{tabular}
\end{ruledtabular}
\end{table*}
\subsection{Symmetry representation in the Fock state basis}
Let us suppose that the symmetry representation
matrix $U(g)$ is given for the electronic operators:
\begin{align}
\hat{g}c_{\vec{R}\tau;\mu}^{\dagger}\hat{g}^{-1}= & \sum_{\nu}c_{\vec{T}\rho;\nu}^{\dagger}[U(g)]_{\nu,\mu},\label{eq:d_transformation}
\end{align}
with $\vec{T}$ and $\rho$ determined by $\vec{T}+\vec{r}_{\rho}=g(\vec{R}+\vec{r}_{\tau})$.

A Fock state $|n_{1},n_{2},\cdots,n_{M}\rangle$ with $N$ occupied
states, where $M$ is the total number of electronic states and $\sum_{l=1}^{M}n_{l}=N$
is written as
\begin{align}
|i_{1},i_{2},\cdots,i_{N}\rangle\equiv & |n_{i_{1}}=1,n_{i_{2}}=1,\cdots,n_{i_{N}}=1\rangle\nonumber \\
= & c_{i_{1}}^{\dagger}c_{i_{2}}^{\dagger}\cdots c_{i_{N}}^{\dagger}|0\rangle,
\end{align}
with $1\le i_{1}<i_{2}<\cdots<i_{N}\le M$. For conciseness, we introduce
a shorthand symbol $I$ representing the ordered set of the integers
$(i_{1},i_{2},\cdots,i_{N})$.

Given the matrix $U(g)$ for the single electron states, the representation
matrix for $\hat{g}$ on the $N$-particle Fock states is
obtained as\begin{widetext}
\begin{align}
\hat{g}|i_{1},i_{2},\cdots,i_{N}\rangle= & \hat{g}c_{i_{1}}^{\dagger}c_{i_{2}}^{\dagger}\cdots c_{i_{N}}^{\dagger}\hat{g}^{-1}|0\rangle\nonumber \\
= & \sum_{k_{1},k_{2},\cdots,k_{N}}c_{k_{1}}^{\dagger}c_{k_{2}}^{\dagger}\cdots c_{k_{N}}^{\dagger}|0\rangle[U(g)]_{k_{1}i_{1}}[U(g)]_{k_{2}i_{2}}\cdots[U(g)]_{k_{N}i_{N}}\nonumber \\
= & c_{j_{1}}^{\dagger}c_{j_{2}}^{\dagger}\cdots c_{j_{N}}^{\dagger}|0\rangle\times\sum_{k_{1},k_{2},\cdots,k_{N}}\text{sgn}\bigg(\begin{matrix}j_{1} & j_{2} & \cdots & j_{N}\\
k_{1} & k_{2} & \cdots & k_{N}
\end{matrix}\bigg)[U(g)]_{k_{1}i_{1}}[U(g)]_{k_{2}i_{2}}\cdots[U(g)]_{k_{N}i_{N}}\nonumber \\
= & |j_{1},j_{2},\cdots,j_{N}\rangle\det\big[[U(g)]_{J,I}\big],
\end{align}
\end{widetext}where the lattice point label $\vec{R}$, the sublattice
label $\tau$, and the spin/orbital label are all absorbed into the
indices $i_{m}$ and $j_{m}$ such that $[U(g)]_{ki}=0$ if the
real-space location of the state indicated by $k$ does not coincide
with the map of the real-space location of the state $i$ by the symmetry
$g$. The result can be compactly written as
\begin{equation}
\hat{g}|I\rangle=|J\rangle\det\big[[U(g)]_{J,I}\big],
\end{equation}
with $I=(i_{1},i_{2},\cdots,i_{N})$ and $J=(j_{1},j_{2},\cdots,j_{N})$.
Here, $[U(g)]_{J,I}$ is a submatrix of $U(g)$
constructed by using the $j_{1},j_{2},\cdots,j_{N}$th rows and $i_{1},i_{2},\cdots,i_{N}$th
columns. In going from the second to the third line, the set of integers
$\{j_{1},j_{2},\cdots,j_{N}\}$ is identified with the ordered set $J$, and $\text{sgn}\bigg(\begin{matrix}j_{1} & j_{2} & \cdots & j_{N}\\
k_{1} & k_{2} & \cdots & k_{N}
\end{matrix}\bigg)$ is the sign of the permutation generated while rearranging the fermion
operators according to the ordering of the electronic states. Note
that $\det\big[[U(g)]_{J,I}\big]\neq\det\big[[U(g)]_{I,J}\big]$
and $\det\big[[U(g)]_{J,I}\big]\neq\det\big[[U(g)]_{I,J}\big]^{*}$
in general.

\subsection{Symmetry transformation of slave-boson operators}

\label{App:Sym_SlaveBoson}The symmetry transformation rule of the
bosonic operators can be found by using the definition of a basis
state $|\vec{R}\tau;\underline{X}\rangle$ in the extended Hilbert
space,
\begin{equation}
|\vec{R}\tau;\underline{X}\rangle=\frac{1}{D_{X}}\sum_{n}\phi_{\vec{R}\tau;Xn}^{\dagger}|0\rangle\otimes|\vec{R}\tau;n\rangle_{f},
\end{equation}
and by supposing that both $|\vec{R}\tau;\underline{X}\rangle$ and $|\vec{R}\tau;n\rangle_{f}$
transform under the symmetry in the same way as the physical
states $|\vec{R}\tau,X\rangle$ and $|\vec{R}\tau,n\rangle$. Here,
$D_{X}^{2}=M!/\{N!(M-N)!\}$, where $N$ is the number of particles
in the many-body state $|\vec{R}\tau;\underline{X}\rangle$, which
is equal to that of $|\vec{R}\tau;n\rangle_{f}$. Introducing the
representation matrices $[{\cal D}(g)]_{XY}$ and $[{\cal D}_{f}(g)]_{mn}$
of $\hat{g}$ with respect to the states $|\vec{R}\tau;\underline{X}\rangle$
and $|\vec{R}\tau;n\rangle_{f}$, the action of the symmetry transformation
is expressed as
\[
[{\cal D}(g)]_{YX}|\vec{T}\rho;\underline{Y}\rangle=\frac{\hat{g}\phi_{\vec{R}\tau;Xn}^{\dagger}\hat{g}^{-1}|0\rangle\otimes[{\cal D}_{f}(g)]_{mn}|\vec{T}\rho;m\rangle_{f}}{D_{X}},
\]
with $\vec{T}$ and $\rho$ determined by $\vec{T}+\vec{r}_{\rho}=g(\vec{R}+\vec{r}_{\tau})$.
The summations over $Y$, $m$, and $n$, which run over
the states with particle number $N_{X}=N_{n}$,
are left implicit for conciseness. By replacing
$|\vec{T}\rho;\underline{Y}\rangle$ on the left-hand side of the
above equation with $\sum_{m}\phi_{\vec{T}\rho;Ym}^{\dagger}|0\rangle\otimes|\vec{T}\rho;m\rangle_{f}/D_{Y}$
with $D_{Y}=D_{X}$, we obtain
\begin{align}
\hat{g}\phi_{\vec{R}\tau;Xn}^{\dagger}\hat{g}^{-1} & =\sum_{Y,m}[{\cal D}^{\dagger}(g)]_{XY}^{*}\phi_{\vec{T}\rho;Ym}^{\dagger}[{\cal D}_{f}(g)]_{mn}^{*},\\
\hat{g}\phi_{\vec{R}\tau;Xn}\hat{g}^{-1} & =\sum_{Y,m}[{\cal D}^{\dagger}(g)]_{XY}\phi_{\vec{T}\rho;Ym}[{\cal D}_{f}(g)]_{mn},
\end{align}
and thus the representation matrix of a symmetry $g$ is given by
\begin{equation}
[\Gamma(g)]_{Ym;Xn}\equiv [\mathcal{D}^{\dagger}(g)]_{XY} [\mathcal{D}_f(g)]_{mn}.
\end{equation}

The representation $\Gamma$ can be decomposed into the irreducible
representations appearing at the $\Gamma$ and the $M$ points of
the space group $P4/nmm$. Including both $\Gamma_{6}$ and $\Gamma_{7}$
CEF levels, we obtain
\begin{align}
\Gamma\approx & \Gamma_{{\rm TRS}}\oplus\Gamma_{{\rm TRSB}},\\
\Gamma_{{\rm TRS}}= & \Gamma_{{\rm TRSB}}\nonumber \\
= & 2A_{1g}\oplus2A_{2g}\oplus2B_{1g}\oplus2B_{2g}\oplus4E_{g}\nonumber \\
 & \oplus2A_{1u}\oplus2A_{2u}\oplus2B_{1u}\oplus2B_{2u}\oplus4E_{u}\nonumber \\
 & \oplus4M_{1}\oplus4M_{2}\oplus4M_{3}\oplus4M_{4},
\end{align}
where $\Gamma_{{\rm TRS}}$ and $\Gamma_{{\rm TRSB}}$ denote the
direct sums of the irreducible representations that are even and odd, respectively,
under time reversal. Because the main text works in the $U\to\infty$ limit, the slave-boson amplitudes there vanish outside the vacuum and single-particle sectors; the symmetry
labels for the single-particle slave-boson channels are summarized in
Table~\ref{tab:irrep_grid}.

As an explicit example, the slave-boson operator corresponding to the $O_{x^{2}-y^{2}}$-type
quadrupolar order is obtained as a linear combination of two basis
matrices that transform according to the $B_{1g}$ representation and preserve
time-reversal symmetry,
\begin{equation}
\hat{\Phi}_{B_{1g}}=a\hat{\Phi}_{B_{1g},1}+b\hat{\Phi}_{B_{1g},2},\;(a,b\in\mathbb{R}),
\end{equation}
with
\begin{widetext}
\small
\begin{equation}
\hat{\Phi}_{B_{1g},1}=\left(\begin{array}{c|cccc|cccc|cccc|cccc}
(10|00|00|00|00|00|00|00) & 0 &  &  &  &  &  &  &  &  &  &  & \\
(01|00|00|00|00|00|00|00) &  & 0 &  &  &  &  &  &  &  &  &  & \\
(00|10|00|00|00|00|00|00) & 1 &  & 0 &  &  &  &  &  &  &  &  & \\
(00|01|00|00|00|00|00|00) &  & 1 &  & 0 &  &  &  &  &  &  &  & \\
\hline (00|00|10|00|00|00|00|00) &  &  &  &  & 0 &  &  &  &  &  &  & \\
(00|00|01|00|00|00|00|00) &  &  &  &  &  & 0 &  &  &  &  &  & \\
(00|00|00|10|00|00|00|00) &  &  &  &  & 1 &  & 0 &  &  &  &  & \\
(00|00|00|01|00|00|00|00) &  &  &  &  &  & 1 &  & 0 &  &  &  & \\
\hline (00|00|00|00|10|00|00|00) &  &  &  &  &  &  &  &  & 0 &  &  & \\
(00|00|00|00|01|00|00|00) &  &  &  &  &  &  &  &  &  & 0 &  & \\
(00|00|00|00|00|10|00|00) &  &  &  &  &  &  &  &  & 1 &  & 0 & \\
(00|00|00|00|00|01|00|00) &  &  &  &  &  &  &  &  &  & 1 &  & 0\\
\hline (00|00|00|00|00|00|10|00) &  &  &  &  &  &  &  &  &  &  &  &  & 0\\
(00|00|00|00|00|00|01|00) &  &  &  &  &  &  &  &  &  &  &  &  &  & 0\\
(00|00|00|00|00|00|00|10) &  &  &  &  &  &  &  &  &  &  &  &  & 1 &  & 0\\
(00|00|00|00|00|00|00|01) &  &  &  &  &  &  &  &  &  &  &  &  &  & 1 &  & 0
\end{array}\right),
\end{equation}
and
\begin{equation}
\hat{\Phi}_{B_{1g},2}=\left(\begin{array}{c|cccc|cccc|cccc|cccc}
(10|00|00|00|00|00|00|00) & 0 &  & 1 &  &  &  &  &  &  &  &  & \\
(01|00|00|00|00|00|00|00) &  & 0 &  & 1 &  &  &  &  &  &  &  & \\
(00|10|00|00|00|00|00|00) &  &  & 0 &  &  &  &  &  &  &  &  & \\
(00|01|00|00|00|00|00|00) &  &  &  & 0 &  &  &  &  &  &  &  & \\
\hline (00|00|10|00|00|00|00|00) &  &  &  &  & 0 &  & 1 &  &  &  &  & \\
(00|00|01|00|00|00|00|00) &  &  &  &  &  & 0 &  & 1 &  &  &  & \\
(00|00|00|10|00|00|00|00) &  &  &  &  &  &  & 0 &  &  &  &  & \\
(00|00|00|01|00|00|00|00) &  &  &  &  &  &  &  & 0 &  &  &  & \\
\hline (00|00|00|00|10|00|00|00) &  &  &  &  &  &  &  &  & 0 &  & 1 & \\
(00|00|00|00|01|00|00|00) &  &  &  &  &  &  &  &  &  & 0 &  & 1\\
(00|00|00|00|00|10|00|00) &  &  &  &  &  &  &  &  &  &  & 0 & \\
(00|00|00|00|00|01|00|00) &  &  &  &  &  &  &  &  &  &  &  & 0\\
\hline (00|00|00|00|00|00|10|00) &  &  &  &  &  &  &  &  &  &  &  &  & 0 &  & 1\\
(00|00|00|00|00|00|01|00) &  &  &  &  &  &  &  &  &  &  &  &  &  & 0 &  & 1\\
(00|00|00|00|00|00|00|10) &  &  &  &  &  &  &  &  &  &  &  &  &  &  & 0\\
(00|00|00|00|00|00|00|01) &  &  &  &  &  &  &  &  &  &  &  &  &  &  &  & 0
\end{array}\right).
\end{equation}
\normalsize
The first column specifies the Fock state
\begin{align}
X={}&|n_{1A\Gamma_{7}+},n_{1A\Gamma_{7}-}|n_{1A\Gamma_{6}+},n_{1A\Gamma_{6}-}|n_{1B\Gamma_{7}+},n_{1B\Gamma_{7}-}|n_{1B\Gamma_{6}+},n_{1B\Gamma_{6}-}\nonumber\\
&|n_{2A\Gamma_{7}+},n_{2A\Gamma_{7}-}|n_{2A\Gamma_{6}+},n_{2A\Gamma_{6}-}|n_{2B\Gamma_{7}+},n_{2B\Gamma_{7}-}|n_{2B\Gamma_{6}+},n_{2B\Gamma_{6}-}\rangle
\end{align}
where $\eta\tau\in\{1A,1B,2A,2B\}$ labels the four sites of the magnetic unit cell, corresponding to each row, while the Fock states $Y$ associated with
the columns follow the same ordering. Thus, each entry represents
the coefficient of the slave-boson operator $\phi_{XY}$. \end{widetext}

\end{document}